# Manifestation of the Verwey Transition in the Tunneling Spectra of Magnetite Nanocrystals

Pankaj Poddar[1], Tcipi Fried[1], Gil Markovich[1](∗), Amos Sharoni[2], David Katz[2], Tommer Wizansky[2], and Oded Millo[2](∗∗)

[1]*School of Chemistry, Raymond and Beverly Sackler Faculty of Exact Sciences, Tel Aviv University, Tel Aviv 69978, Israel*

[2]*Racah Institute of Physics and the Center for Nanoscience and Nanotechnology, The Hebrew University of Jerusalem, Jerusalem 91904, Israel*

**ABSTRACT**

Tunneling transport measurements performed on single particles and on arrays of $Fe_3O_4$ (magnetite) nanocrystals provide strong evidence for the existence of the Verwey metal-insulator transition at the nanoscale. The resistance measurements on nanocrystal arrays show an abrupt increase of the resistance around 100 K, consistent with the Verwey transition, while the current-voltage characteristics exhibit a sharp transition from an insulator gap to a peak structure around zero bias voltage. The tunneling spectra obtained on isolated particles using a Scanning Tunneling Microscope reveal an insulator-like gap structure in the density of states below the transition temperature that gradually disappeared with increasing temperature, transforming to a small peak structure at the Fermi energy. These data provide insight into the roles played by long- and short-range charge ordering in the Verwey transition.

(∗) E-mail:  gilmar@post.tau.ac.il, (∗∗)milode@vms.huji.ac.il

The Verwey metal-insulator transition observed in magnetite ($Fe_3O_4$) has continuously attracted interest since its discovery more than 60 years ago [1,2]. Magnetite is a relatively good conductor at room temperature and was shown to be a highly spin polarized conductor [3]. On cooling below 120 K its conductivity sharply drops by two orders of magnitude. It was described as a first-order metal-insulator transition accompanied by a structural phase transition where the cubic symmetry of the $Fe_3O_4$ crystal is broken by a small lattice distortion [4]. While the exact nature of the transition is still under controversy, it is understood that the driving force for this phenomenon is the strong electron-electron and electron-lattice interactions in the system.

Special attention has been focused on the roles played by long-range and short-range charge ordering in driving the Verwey transition. The former is believed to exist below the transition temperature, $T_V$, and the latter sustains well above it. The long-range order manifests itself by opening a gap in the electronic density of states (DOS) around the Fermi level ($E_F$). This gap was detected by photoemission [5,6], optical [7], and tunneling [8] spectroscopies. The effect of short-range ordering on the DOS is not as clear. However, recent photoemission experiments [5,6] suggest that a reduced gap in the DOS, attributed to short-range ordering, still exists well above the transition temperature. It should be noted, however, that all the above experiments were performed on bulk systems.

In view of the apparent importance of long-range charge ordering in determining the electronic properties below $T_v$, intriguing questions arise: How small could the system be for the Verwey transition be observed? How would this transition and the corresponding electronic structure evolve with particle size? Partial answers to these questions are given by the recent work of Poddar et al. where a sharp Verwey transition in arrays of $Fe_3O_4$ (magnetite) nanoparticles of average size around 5.5 nm was observed [9]. However, it is not clear whether the observed transition pertained to the single nanocrystal (NC) or was a collective array effect. In this Letter we present scanning tunneling spectroscopy results on single nanoparticles, providing evidence that the transition occurs in isolated magnetite NCs of diameters measuring only several unit cells. The tunneling spectra show a gap structure in the DOS below $T_v$, which gradually transforms into a peak around $E_F$. These data are in contrast to measurements performed on bulk systems, where a gap was found to sustain well above $T_v$, raising questions about the effects of long- and short-range charge ordering on the DOS.

The details of magnetite NCs array preparation are described elsewhere [10]. After synthesis, the NCs were coated with oleic acid, dispersed in non-polar organic solvents and separated by employing size selective precipitation to produce samples of NCs within the 3-10 nm size range having about 20% size distribution. Transmission electron microscopy (TEM) was used for size determination as well as to confirm the formation of NC monolayers. The monolayers were produced by depositing the hydrophobic magnetite nanocrystals from heptane solution at the air-water interface, and then compressing the NCs to form two-dimensional close-packed arrays. For the fabrication of the junctions, arrays of gold lines (width = 80μm) and a thickness of 100 nm were deposited by thermal evaporation through a shadow mask on a doped Si substrate with a 100 Å thick oxide layer. Consequently, five monolayers of $Fe_3O_4$ NCs were transferred from the air-water interface onto the patterned substrates using the Langmuir-Schafer technique. A second array of gold lines perpendicular to the bottom electrodes was then deposited on top of the films while the substrate was maintained at 20 °C. The electrical contacts were formed by fixing Cu wires to pads on the gold electrodes, using silver paste [9].

For the scanning tunneling microscopy (STM) measurements, the bare magnetite particles were spin-coated on gold films deposited on mica substrates and then dipped in oleic acid solution. The fatty acid molecules mechanically stabilized the NCs and served as a protecting layer against their oxidation. All this procedure was performed in an inert atmosphere. The low temperature tunneling spectra were obtained using the double barrier tunnel junction configuration, by positioning the STM tip above single NCs [11,12]. The tip was retracted as far as possible from the NC in order to form a highly asymmetric tunneling configuration and thus reduce single electron charging effects [12,13]. This was facilitated by the preparation procedure that ensured a relatively large capacitance and small resistance (large tunneling rate) for the substrate-NC tunnel junction. Nevertheless, about 60% of the NCs did show charging effects that dominated the tunneling characteristics, thus masking the intrinsic NC density of states. Eleven out of fourteen NCs that did not exhibit charging effects (see below) showed a clear opening of a gap in the DOS around $T_v$, associated with the Verwey transition, thus forming a reasonable statistical ensemble for our investigation.

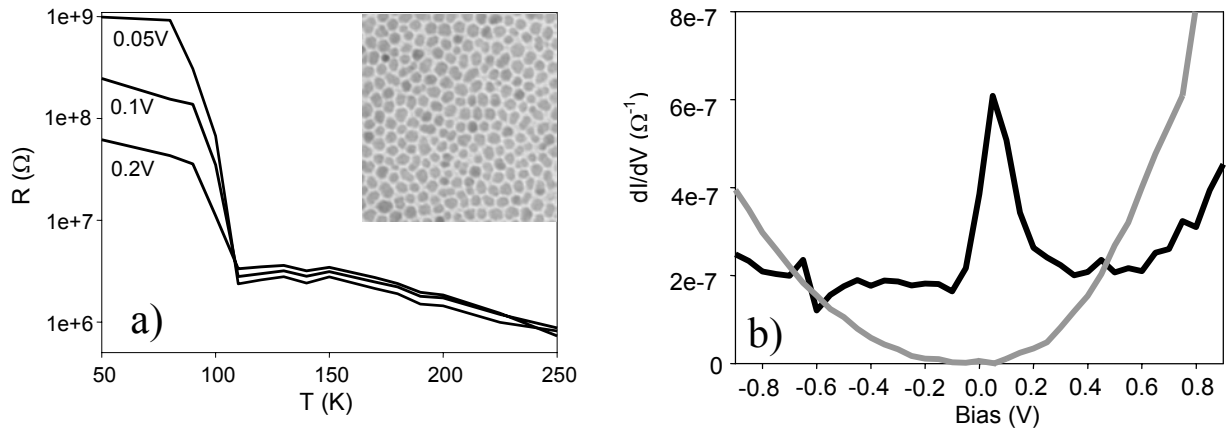

Fig. 1. (a) Resistance of a magnetite nanocrystal array as function of temperature for three bias voltage values. Inset: TEM image of a magnetite NC monolayer with average particle size of 5.5 nm. (b) Differential conductance vs. bias voltage measured on the array above (black) and below (gray) the Verwey transition temperature.

The temperature dependence of the two-terminal resistance of a macroscopic tunnel junction is shown in Figure 1a. The sharp resistance jump observed at 100±5 K clearly manifests the first-order Verwey transition. The typical transition temperature measured for the arrays is smaller than $T_v$ observed for high-quality bulk samples (~120 K). This may be due to surface defects, stoichiometry deviation, or finite size effects. This temperature is independent of the bias voltage used in the measurement but the magnitude of the resistance jump decreases with increasing bias due to non-linearity. In Figure 1b we present two dI/dV vs. V tunneling spectra, one measured above (black) and the other below (gray) $T_v$, showing significantly different behaviors. A pronounced change in the tunneling conductance curves takes place on cooling through $T_v$. The sharp peak around zero bias ($E_F$) seen right above $T_v$ disappears abruptly, and a region of low conductance sets-in below $T_v$. This effect was reversible and reproducible through several cooling-heating cycles [9].

As discussed previously [9], the macroscopic tunnel junctions are perceived as consisting of parallel percolating current paths, where each path may include several inter-particle junctions. Hence, the details of the current-voltage curves were determined not only by the single-particle electronic structure but also by the $RC$ constants of the various inter-particle tunneling junctions along the current paths. However, since the $RC$ characteristics of the junctions were not sensitive to temperature, we believe that the abrupt change in the tunneling spectra is a manifestation of the metal-insulator transition in the single particle. In particular, it appears that in the metallic phase (above 100 K), the conductance peak near the Fermi level reflects the narrow conduction band of magnetite, as previously observed on bulk samples [14]. Below $T_v$ the region of low conductance around zero bias results from opening a gap in the DOS around $E_F$ of the NCs. Moreover, magnetoresistance measurements reported elsewhere [9] clearly indicate that electrical transport through this NC array is dominated by the intrinsic electronic structure of the NCs rather than by extraneous charging effects. The gap that opened in the DOS is therefore attributed to the Verwey transition, but its width may be affected also by the inter-particle junction parameters. Tunneling spectroscopy measurements on single particles, as enabled by STM, may therefore yield further information on the evolution of the gap in single NCs around $T_v$, as discussed below.

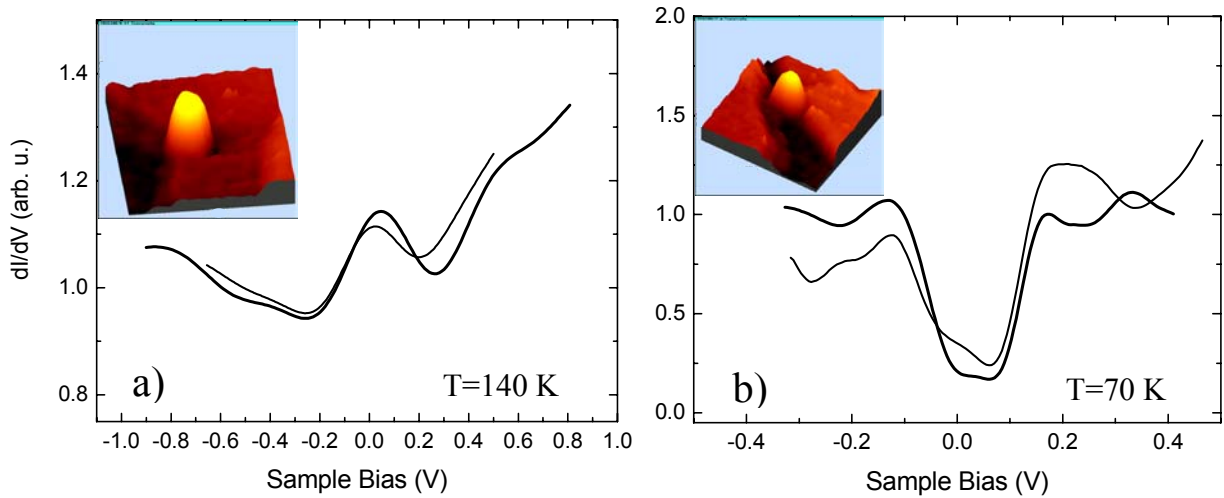

Fig. 2. Tunneling spectra taken on isolated magnetite NCs above (a) and below (b) the Verwey transition temperature. The two curves in each frame were taken over the same NC with the same conditions, showing the variance in the data. The insets present STM images of the corresponding NCs (image sizes are: (a) 26x30 nm, (b) 52x40 nm).

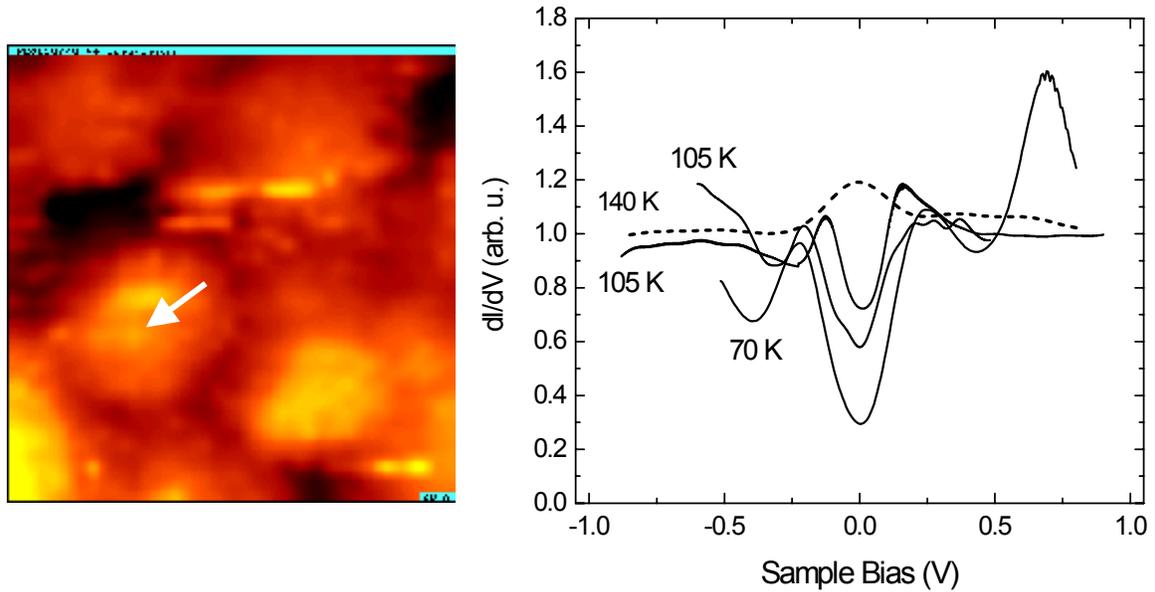

Fig. 3. 20x20 nm² STM image showing a cluster of magnetite NCs (left) together with a series of tunneling spectra taken nominally at the position marked by the arrow at various temperatures, as indicated (right).

Fig. 2 presents tunneling spectra measured on single magnetite NCs, about 8 nm in diameter (the tip-nanocrystal convolution does not allow for an accurate determination of the NC size, which may be somewhat overestimated). The curves in Fig. 2(a) were measured at 140 K, well above $T_v$, whereas those in Fig. 2(b) were acquired at 70 K. Consistent with the macroscopic tunnel junction measurements, a peak in the DOS is observed above the Verwey transition while a gap develops below the transition. The reproducibility of the data can be appreciated from two typical curves plotted for each particle: the main features, namely, the peak and gap structures, are reproducible, but the background may fluctuate. Upon changing the temperature, the tip and the NCs usually drifted apart to a distance beyond the available scanning range of our STM and thus we were not able to measure, in most cases, the same NC well above and well below $T_v$.

The measured gap value in Fig. 2(b) is about 250 meV, which is significantly larger than previously measured for bulk samples using photoemission, ~140 meV (assuming a symmetric gap around the Fermi level) [5,6] or by STM, ~200 meV [8]. This apparent gap broadening may be due to either the effect of voltage division, inherent to the double barrier tunnel junction configuration [12,13] or to a finite size effect that is not understood at this point. This gap, however, could not be washed out by changing the STM bias and current settings over a wide range, and no signature of an equidistant peak structure at higher voltages was ever observed in the conductance spectra measured on this NC. This rules out the possibility that the observed gap is associated with single electron charging effects (i.e., the Coulomb Blockade and Coulomb Staircase) [11,15]. As noted above, eleven NCs exhibited a similar behavior, where a gap, which is clearly intrinsic and *not* associated with single electron charging, opened in the DOS below $T_v$. The apparent gaps, well below $T_v$, varied between 250 to 350 meV (mostly up to 300 meV), reflecting variations in the quality (mainly stoichiometry and surface conditions) as well as the tunnel-junction parameters from one NC to another (the latter determining the amount of gap broadening). The width of the zero bias peak above $T_v$ was typically 400 meV, somewhat larger than the value obtained using the macroscopic tunnel junction, ~ 300 meV (see Fig. 1).

Figure 3 depicts the evolution of the DOS around $E_F$ with temperature. Here, the tunneling spectra were acquired on a small cluster of (single layer) NCs, nominally at the position marked by the white arrow. One can see that as the temperature is increased, the gap reduces, and the DOS at the Fermi energy gradually increases. The two spectra at 105 K may have been obtained (due to drift of the tip) at somewhat different positions, resulting in a different tunneling DOS. In these measurements it appears that a clear gap still exists somewhat above 100 K (the transition temperature measured for the NC arrays) and the peak structure emerged only at a higher temperature, in contrast to the more abrupt change observed for the arrays, as discussed above. This may reflect the particle-to-particle fluctuations (e.g., in composition or surface condition), to which the STM is sensitive, as opposed to the macroscopic tunnel junction where the ensemble properties are measured. We recall that some of the NCs did not exhibit any transition behavior in the STM spectra.

Interestingly, the observation of a prominent gap structure in the DOS below $T_v$ indicates that a magnetite NC having a dimension of only several unit cells is sufficient to support long-range charge order as in bulk systems. However, the suppression of the gap in bulk crystals, as observed in photoemission and optical spectroscopy experiments [5,6,7], occurred in a much more gradual way as a function of temperature as compared to our observations for the $Fe_3O_4$ NCs. For example, in the former experiments a gap was detected up to nearly room temperature. Since the gap in the DOS above $T_v$ was attributed [5,6,7] to the short-range charge ordering, it is expected that this gap would exist also in the nanocrystals, and it is therefore puzzling why it was not detected in our tunneling spectra. Within the framework of the long- to short-range charge ordering picture this puzzle emphasizes the need for a better understanding of how short-range order manifests itself in the tunneling (single particle) size dependent DOS of magnetite.

In summary, the combination of macroscopic tunnel junctions and scanning tunneling spectroscopy is demonstrated to be an effective tool for studying the electronic properties of $Fe_3O_4$ nanocrystals in the vicinity of the Verwey transition. In particular, the STM measurements clearly show that the sharp resistive transition observed for NC arrays is associated with, and probably governed by, an opening of a gap in the electronic DOS around $E_F$ in single NCs. This result suggests that the effect of long-range order, that is believed to dominate the electronic properties at $T<T_V$, is significant even in nanometer size magnetite particles. More detailed size dependent experiments are still needed to resolve the finite size effects on the transition temperature and the electronic structure of the magnetite system.

***

This work was supported in parts by the Israel Science Foundation, the Israel Ministry of Science, and the DIP Foundation.